\documentclass[12pt]{article}
\usepackage{times}
\usepackage{graphicx}
\usepackage{epsfig}
\usepackage[koi8-r]{inputenc}
\usepackage[russian]{babel}
\begin{document}
\pagestyle{plain}
\begin{titlepage}
\flushright{IHEP 2008-27}
\vspace*{0.15cm}
\flushright{OEF}
\vspace*{0.15cm}
\begin{center}
{\Large \bf Observation of the destructive interference in \\ 
the radiative kaon decay K$^- \rightarrow \mu^- \bar \nu \gamma$ \\}
\vspace*{0.3cm} 
{\bf O.G.~Tchikilev, S.A.~Akimenko, G.I.~Britvich, A.P.~Filin,
 A.S.~Konstantinov, I.Y.~Korolkov, V.M.~Leontiev, V.F.~Obraztsov, 
 V.A.~Polyakov, V.I.~Romanovsky, V.K.~Semenov, V.A.~Uvarov,
 O.P.Yushchenko}
\vskip 0.15cm 
{\large \bf $Institute~ for~ High~ Energy~ Physics,~ Protvino,~ Russia$}
\vskip 0.35cm 
{\bf V.N.~Bolotov, V.A.~Duk, A.I.~Makarov, A.A.~Khudiakov, V.P.~Novikov, 
A.Yu.~Polyarush}
\vskip 0.15cm  
{\large \bf $Institute~ for~ Nuclear~ Research,~ Moscow,~ Russia$}       
\end{center}
\end{titlepage}
\vspace*{0.3cm}
\begin{center}
Abstract
\end{center}
\vspace*{0.15cm}

 Using data collected with the ``ISTRA+'' spectrometer at
   U70 proton synchrotron of IHEP, we report the first measurement of 
 the destructive interference in the radiative kaon decay 
 K$^{-} \rightarrow \mu^{-}\overline{\nu}\gamma $.  
 We find the difference of the vector and axial form factors
  $F_V - F_A = 0.126 \pm 0.027($stat$) \pm 0.043($syst$)$. The measured value 
  is two standard deviations above the O(p$^4$) ChPT prediction 
  equal to 0.055. Inclusion of exotic tensor interaction gives 
  $F_V-F_A=0.144 \pm 0.044($stat$) \pm 0.035($syst$) $  and
 $ F_T = -0.0079 \pm 0.0113($stat$) \pm 0.0073($syst$) $,
  i.e. $-0.03 < F_T < 0.01$
  at $90\%$ CL,
  consistent both with zero and with recent theoretical prediction
 equal to $|F_T| =0.022$. 
\renewcommand{\refname}{References}
\renewcommand{\figurename}{Figure}
\renewcommand{\tablename}{Table}
\vskip 0.6cm
\section{ Introduction}

  The decay $K^-\rightarrow \mu^-\overline{\nu}\gamma$ proceeds
  via two distinct mechanisms: the internal Bremsstrahlung (IB) with the photon
  emitted by the kaon or the muon, and the structure-dependent(SD) decay
  involving emission of the photon from intermediate states. SD is
  sensitive to the electroweak structure of the kaon and
   allows for good test of theories describing hadron interactions and
  decays, like Chiral Perturbation Theory (ChPT)\cite{chpt,ref1}. This
  decay is also a good place for  searches for possible tensor interactions,
  see theoretical~\cite{tens1,tens2,tens3,tens4,tens5} and 
  experimental~\cite{isb1,pib1,pib2,pib3} papers, devoted mainly to the
  $\pi \rightarrow e\nu\gamma$ decay.   
  The differential probability of the decay  can be
  written in terms of $x=\frac{2E_{\gamma}}{M_K}$ and
  $y=\frac{2E_{\mu}}{M_K}$ (where $M_K$ is the kaon mass and $E_{\gamma}$
  and $E_{\mu}$ are the photon and muon energies in the kaon rest frame).
  In the most general case, which includes hypothetical tensor term it
  reads:
\begin{eqnarray*}
 \frac{d \Gamma }{dxdy}= A_{IB}f_{IB}(x,y)+A_{SD}[(F_V+F_A)^2
 f_{SD^+}(x,y)
  +(F_V-F_A)^2 f_{SD^-}(x,y)] \\
  -A_{INT}[(F_V+F_A)f_{INT+}(x,y)+
  (F_V-F_A)f_{INT-}(x,y)] 
  +A_{T}F_T^2f_{T}(x,y) \\
  -A_{IBT}F_T f_{IBT}(x,y) 
  -A_{SDT}F_T (F_V-F_A)f_{SDT}(x,y).
\end{eqnarray*}
\begin{equation}
 f_{IB}(x,y)=[\frac{1-y+r}{x^2(x+y-1-r)}][x^2+2(1-x)(1-r)-
  \frac{2xr(1-r)}{x+y-1-r}],
\end{equation}
\begin{equation}
f_{SD^+}(x,y)=[x+y-1-r][(x+y-1)(1-x)-r],
\end{equation}
\begin{equation}      
 f_{SD^-}(x,y)=[1-y+r][(1-x)(1-y)+r],
\end{equation}
\begin{equation}
f_{INT+}(x,y)=[\frac{1-y+r}{x(x+y-1-r)}][(1-x)(1-x-y)+r],
\end{equation}
\begin{equation}
f_{INT-}(x,y)=[\frac{1-y+r}{x(x+y-1-r)}][x^2-(1-x)(1-x-y)-r],
\end{equation}
\begin{equation}
f_T(x,y) =(x+y-1-r)(1+r-y)
\end{equation}
\begin{equation}
f_{IBT}=(1+r-\frac{x+y-1-r}{x}-\frac{rx}{x+y-1-r})
\end{equation}
\begin{equation}
f_{SDT}= x(1+r-y)
\end{equation}
where $r=(\frac{M_{\mu}}{M_K})^2$ with $M_{\mu}$ being the muon mass and  
\begin{equation}
A_{SD} = \Gamma_{K_{\mu 2}}\frac{\alpha}{8\pi}\frac{1}{r(1-r)^2}
 [\frac{M_K}{F_K}]^2,
\end{equation}
$A_{IB} = 4r(\frac{F_K}{M_K})^2 A_{SD}$ ,
$A_{INT} =4r(\frac{F_K}{M_K}) A_{SD}$ ,
 $A_T = 4 A_{SD}$, $A_{SDT} = 4\sqrt{r} A_{SD}$ and
$ A_{IBT} = 8 \sqrt{r} (\frac{F_K}{M_K}) A_{SD}$.

 In these formulas $F_V$ and $F_A$ are the vector and axial
 form factors, $F_T$ is the tensor form factor. We use the 
 prescription~\cite{tens1,gabr1} for the tensor interaction. 
 $\alpha$ is the fine structure constant,
 F$_K$ is the charged kaon decay constant ($155.5 \pm 0.2
 \pm 0.8 \pm 0.2$~MeV~\cite{pdg2}, and 
 $\Gamma_{K_{\mu 2}}$ is the width of the
 K$_{\mu2}$ decay. 
As in the paper\cite{bnl}
   a minus sign precedes the interference terms, thus changing the
   sign of the form factors.  
Without this change in sign the formulas coincide with the ones
given in \cite{tens1,gabr1}.   
   SD$^+$ and SD$^-$ refer to the terms with different photon polarizations and
   do not mutually interfere. Their interference with IB leads
   to the terms labeled INT$^+$ and INT$^-$. The  term SDT
   refers to SD$^-$T interference, the interference with SD$^+$
    is possible only in rather exotic
   models~\cite{tens2,tens3,tens4}. The interference between
   the tensor  
   and inner bremsstrahlung amplitudes is labeled as IBT.
    
   The x vs y plots
   for different terms are illustrated in Figs.1 and 2. The 
   quadrangle
   area  in these figures is restricted by the
   condition $x_b < x < x_b+0.1$ for the $y$ interval 0.49---1.0 ,
   where  $x_b = 1.0- 0.5(y+\sqrt{y^2-4r})$
   is the border of the Dalitz plot. This area, characterized by
   considerable contribution from INT$^-$,SD$^-$, IBT and SDT terms, and
   favourable background conditions 
     is used in our analysis.
       
   Generally form factors can depend on $q^2=(p_K-p_{\gamma})^2
   =M_K^2(1-x)$. In our analysis we assume either constant form
   factors or
   the  phenomenological
   dependence
   for the vector and axial form factors
   as in \cite{bnl}: $F_V(q^2)=F_V(0)/(1-q^2/M_V^2)$ and
   $F_A(q^2)=F_A(0)/(1-q^2/M_A^2)$ with $M_V=0.870$~GeV and
   $M_A=1.270$~GeV. The theoretical situation with $q^2$ dependence
   of the form factors is controversial. For $O(p^4)$ ChPT 
   calculations~\cite{chpt} there is no $q^2$ dependence, 
   $O(p^6)$ ChPT calculations~\cite{geng1,geng2} predict nearly
   linear dependence, similar to the phenomenological one. At the
   same time calculations~\cite{geng1,geng2}
    in the framework of the light-front quark
   model(LFQM) give quite different dependence with form factors
   going to zero with $q^2$ increase.
   
   We assume also the $q^2$ independence of the tensor form 
   factor $F_T$.
   
   The absolute value of the sum of the
   vector and axial form factors is known with
   high precision: $|F_V+F_A| = 0.155 \pm 0.008$\cite{bnl}, whereas
   the difference $F_V-F_A$ is still poorly known.
   The latest measurent~\cite{bnl} gives for $F_V-F_A$ only
   the 90\% confidence level: 
    $-0.04<F_V-F_A<0.24$, whereas the $O(p^4)$ ChPT prediction
    is equal to 0.055~\cite{chpt,ref1}. 
   
   The situation with the tensor contribution is controversial. The measurements
  have been done mainly for pion decays\cite{isb1,pib1,pib2,pib3}, The observation
  of the ISTRA experiment\cite{isb1}  was confirmed 
  in the preliminary result by the
  PIBETA Collaboration\cite{pib1,pib2} having statistically significant
  deviation from the standard model without tensor term. Latest PIBETA
  data\cite{pib3} have however eliminated the evidence for the tensor term
  and give the limit: $-0.00052< F_T < 0.0004$ at the 90\% CL. The searches
  for tensor terms are done also in the studies of the
   angular distributions in the
   Gamow-Teller $\beta$
  decays of the spin polarized atoms, see for example \cite{atom} and the 
  references therein. These studies are still less restrictive than 
  $\pi_{e2\gamma}$ data.
   
   It is of interest to note that in some theoretical models~\cite{gabr1}
   $F_T$ for kaon decay is larger by a factor of $1/\tan(\theta_C) \sim 4.4$
   as compared with pion decay. Here $\theta_C$ is the Cabibbo angle.
 
\begin{figure}
\epsfig{file=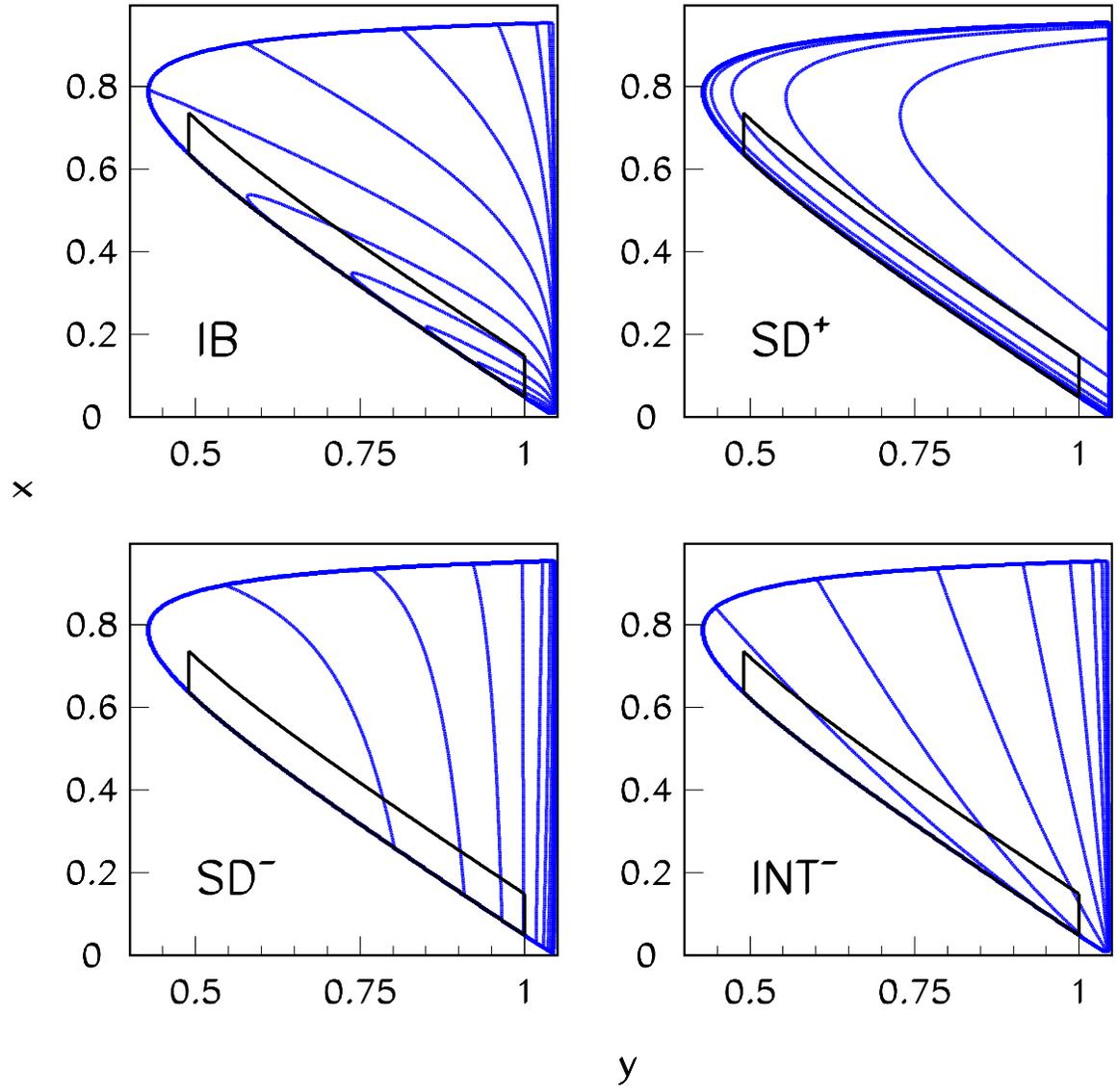,width=17.5cm}
\caption{Dalitz plots for IB, SD$^+$, SD$^-$ and INT$^-$ 
 contributions. The vertical scale  is logarithmic. The 
 quadrangle area shows the region studied. }
\end{figure}   
\begin{figure}
\epsfig{file=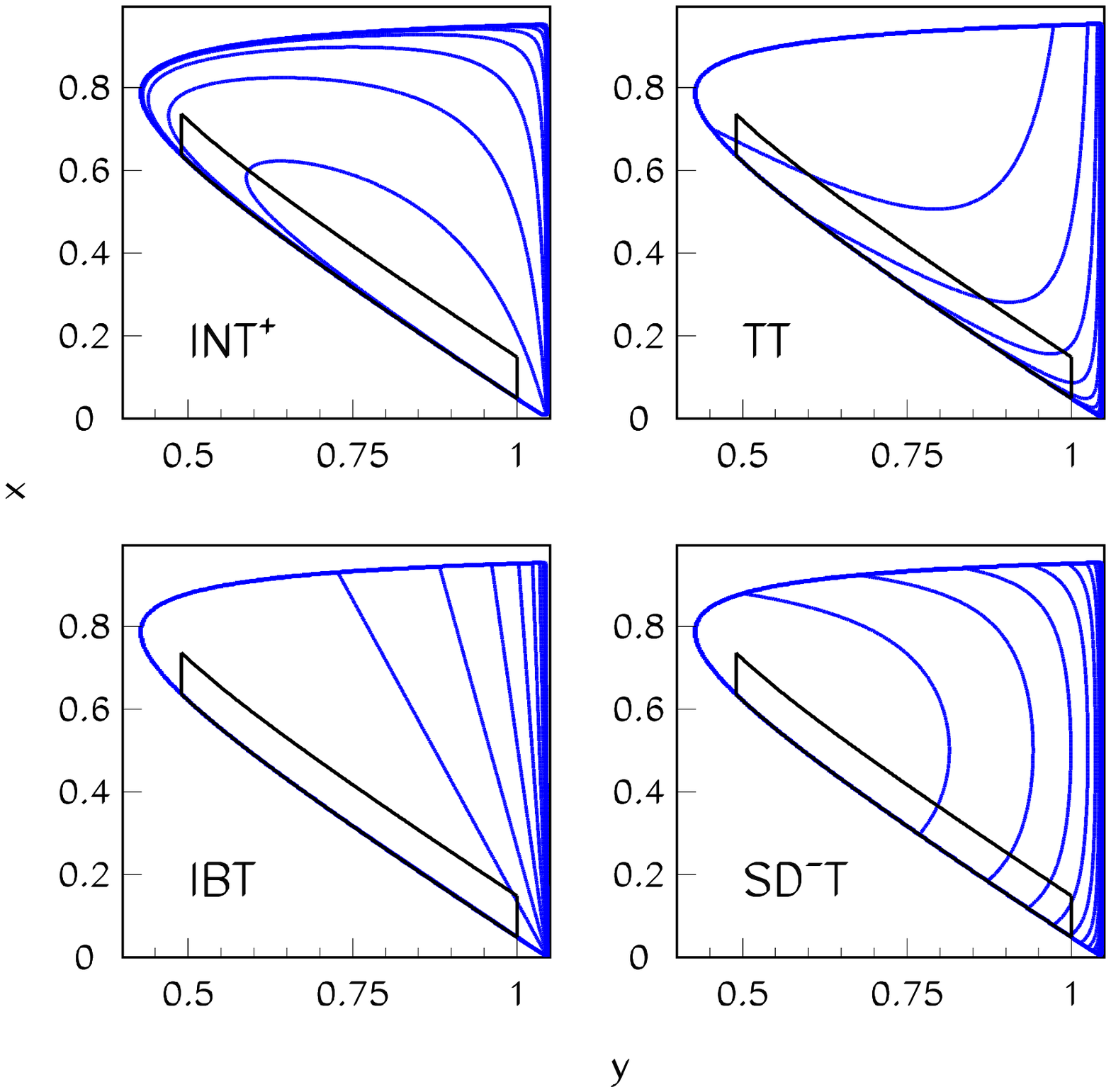,width=17.5cm}
\caption{The same as in Fig.1 for INT$^+$, T, IBT and SD$^-$T terms.
 }
\end{figure}

\section{ Experimental setup and event selection}

The experiment is performed at the IHEP 70 GeV proton synchrotron U70.
The    ISTRA+ spectrometer has been described in detail in 
 papers on $K_{e3}$  \cite{ref2,ref2n}, $K_{\mu 3}$ \cite{ref3,ref3n} 
  and $\pi^-\pi^{\circ}\pi^{\circ}$ decays \cite{ref4}.
Here we recall briefly  the characteristics relevant to our analysis. 
 The  ISTRA+ setup is located in the negative unseparated 
 secondary beam line 4A of the U70. The beam momentum  is $\sim 25$ GeV/c with 
$\Delta p/p \sim 1.5 \%$. The admixture of $K^{-}$ in the beam is $\sim 3 \%$,
 the beam intensity is $\sim 3~\cdot~10^{6}$ per 1.9 sec U70 spill.

 During the  physics run in November-December 2001 350 million trigger
 events were collected with  high beam intensity. This information
 is complemented by 124~M  Monte Carlo~(MC) events generated using 
 Geant3 \cite{ref5}
 for the dominant $K^-$ decay modes, 100~M of them are the mixture of the dominant
 decay modes with the branchings exceeding 1~\% and
  24~M MC
  events are the radiative $K_{\mu2}$~ decays.
    
 Some information on the data
 processing and reconstruction procedures is given in
 \cite{ref2,ref2n,ref3,ref3n,ref4},
 here we briefly mention the details relevant for  present analysis.
 
 The muon identification (see \cite{ref3,ref3n}) is based 
 on the information from the
  electromagnetic calorimeter SP$_1$ and hadron calorimeter
   HC. The energy deposition in the SP$_1$ is required to be compatible
  with the MIP signal in order to suppress charged pions and electrons. The sum of
  the signals in the HC cells associated with charged track is  required to
  be compatible with the MIP signal.  The muon selection is further enhanced by
  the requirement that the ratio $r_3$
   of the HC energy in last three
  layers to the total HC  energy  exceeds 5~\%. The used cut values are
  the same as  in~\cite{ref3n}.
  
   Events with one reconstructed charged track and one reconstructed
   shower in the calorimeter SP$_1$ are selected. 
     
  A set of cuts is developed  to suppress various  
 backgrounds  and/or to do data cleaning:
 
 0) We select events with
  good charged track having two reconstucted ($x-z$ and $y-z$) projections and the number of
 hits in the matrix hodoscope MH below 3. 
 
 1) Events with the reconstructed vertex inside the interval
  $ 400 < z < 1600$~cm are selected.
  
 2) The measured missing energy 
 $E_{mis}= E_{beam}-E_{\mu}-E_{\gamma}$ is
  required to be above zero.
   
 3)        The events with missing momentum pointing to the
    SP$_1$ working aperture are selected in order to suppress 
 $\pi^-\pi^{\circ}$ background ( $r>10$~cm, here $r$ is the distance between
 the impact point of the missing momentum and the SP$_1$ center in the SP$_1$ transverse 
 plane).
 
 4) We require also the absence 
 of the signal above the threshold in the calorimeter SP$_2$ and
 the guard veto system GS.
 
 5) In order to suppress the remaining K$_{\pi2}$ contribution
  at $y$ around 0.8 we use the cut $\cos (\phi (\mu\gamma)) > -0.95$
  where $\phi ( \mu\gamma)$ is the angle between transverse momenta
  of the muon and the photon.
 
  We look for the signal in the distributions over the effective mass
  M$(\mu^-\gamma\nu)$, where $\nu$ four-momentum is 
  calculated using the
   measured missing momentum and assuming m$_\nu=0$. 
   Effective mass spectra for the quadrangular region in Figs.~1 and~2 are
   shown in Figs.~3,4 and 5 for $y$-interval 0.49---1.00 with
   the step $\delta y=0.03$.        
\begin{figure}
\epsfig{file=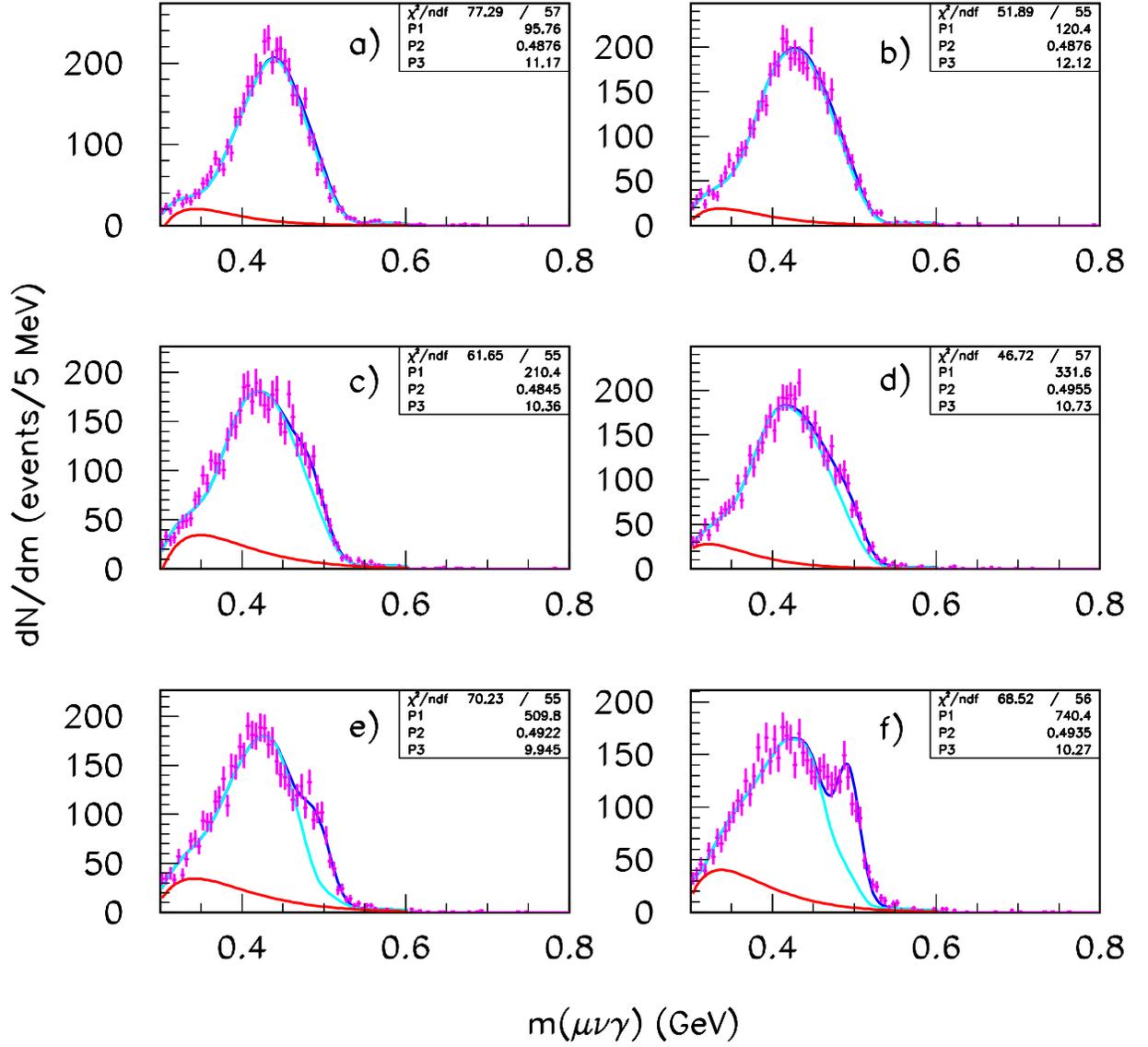,width=17.5cm}
\caption{ Effective mass m$(\mu^-\nu\gamma$) spectra
for the $y$-interval 0.49---0.67 with the step $\delta y=0.03$.}
\end{figure}
\begin{figure}
\epsfig{file=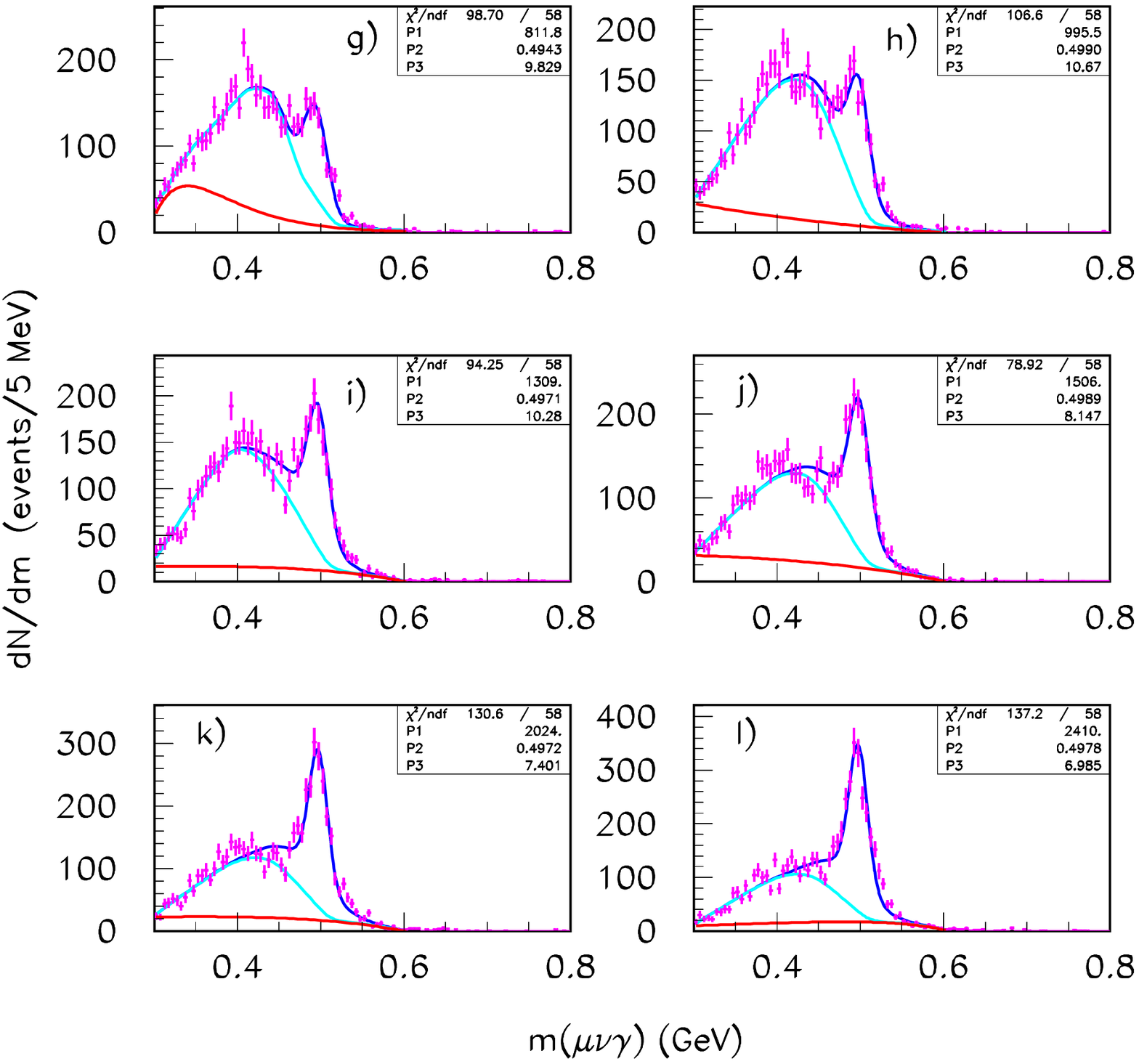,width=17.5cm}
\caption{ Effective mass m$(\mu^-\nu\gamma)$ spectra
 for the $y$-interval 0.67---0.85 with the step $\delta y=0.03$.
 }
\end{figure}
\begin{figure}
\epsfig{file=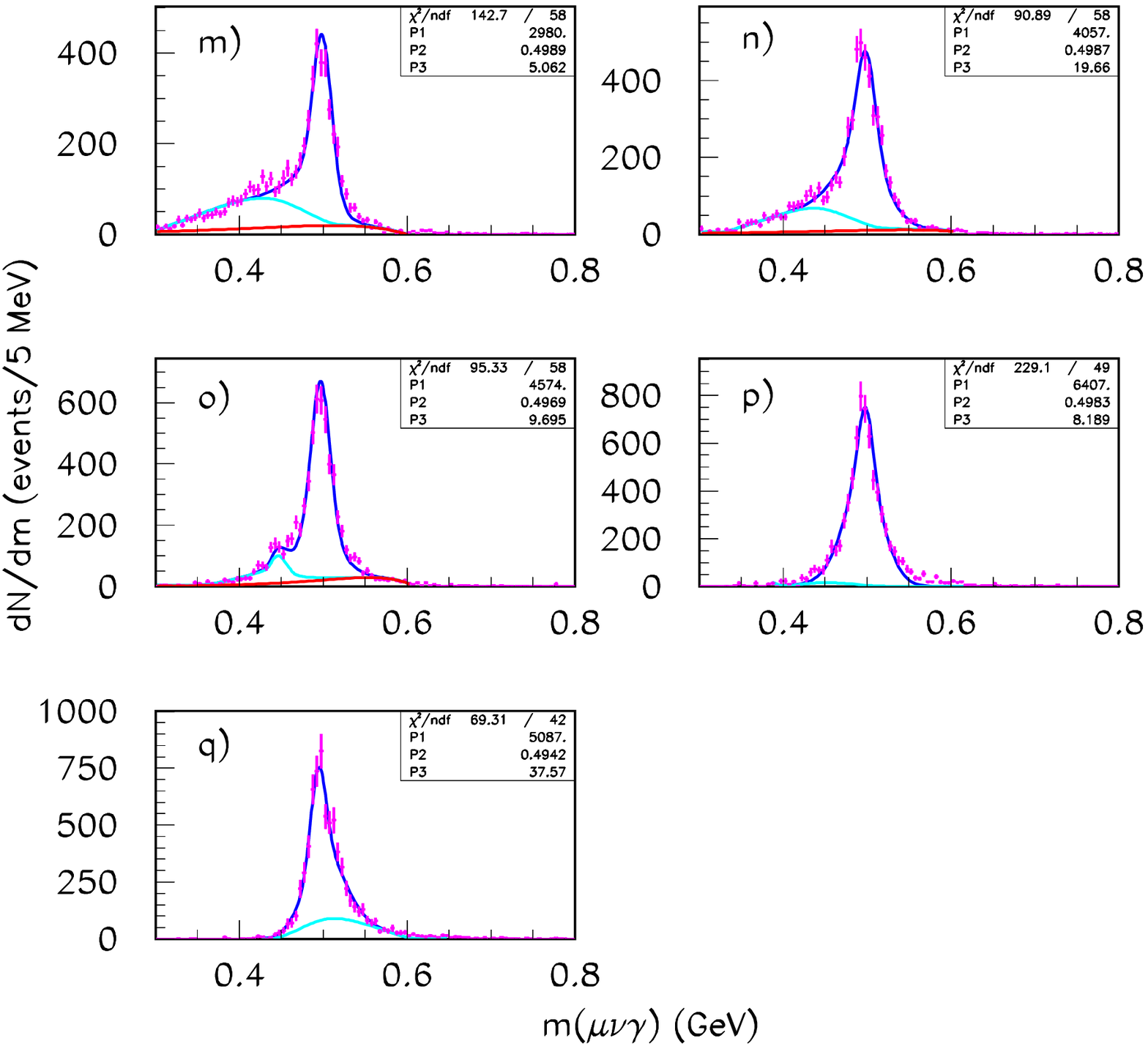,width=17.5cm}
\caption{Effective mass m$(\mu^-\nu\gamma)$ spectra
 for the $y$-interval 0.85-1.00 with the step $\delta y=0.03$.}
\end{figure}
 
 The effective mass spectra have been parametrized by the sum of the signal
 and of the background. The signal form have been found from the
 signal Monte Carlo events parametrized by the sum of two Gaussians. 
 The background have been found using the histogram smoothing of
 the MC background mass spectra
 by the HQUAD routine from the HBOOK package~\cite{hquad}. This
 background does not ideally describe the real data, especially at
 low effective masses.
  This discrepance
 has been taken into account by addition of the term
 \begin{equation}
 (P(4)+P(5)*M)*\exp{(-P(6)*M)}
 \end{equation}
   to the MC background. It should be noted that the contribution of this term
   in the signal regionis rather small.
    
   First parameter of the
 fit gives the number of events in the kaon peak, second --
 the position of the peak, third -- normalization of the MC
 background. 
 
  At small $y$ the signal is rather small, at large $y$, especially
  in the IB region, it dominates over the background. The peak
  at the effective mass 0.43--0.45~GeV, seen in the histogram
   o) is the reflection of the remaining $K_{\pi2}$ background.
  
  The resulting event distribution in the interval $0.49< y<1.00$
  have been parametrized by function constructed using Phase Space
  signal MC events, weighted with corresponding terms (1)-(8) calculated
  using ``true'' MC $x$ and $y$ values
in  the corresponding $\delta y$ intervals. 
  The results of the fit without tensor component are shown
  in Fig.6. Here the first parameter is $F_V+F_A$ fixed at the
  value 0.155 taken from \cite{bnl}, the second parameter is
  $F_V-F_A$, the third parameter is the tensor form factor
   and  the fourth parameter is the normalization
  factor. In fact, the fit results are insensitive to the value
  $F_V+F_A$ since the SD$^+$ and INT$^+$ contributions are
  negligible, see Fig.6.
\begin{figure}
\epsfig{file=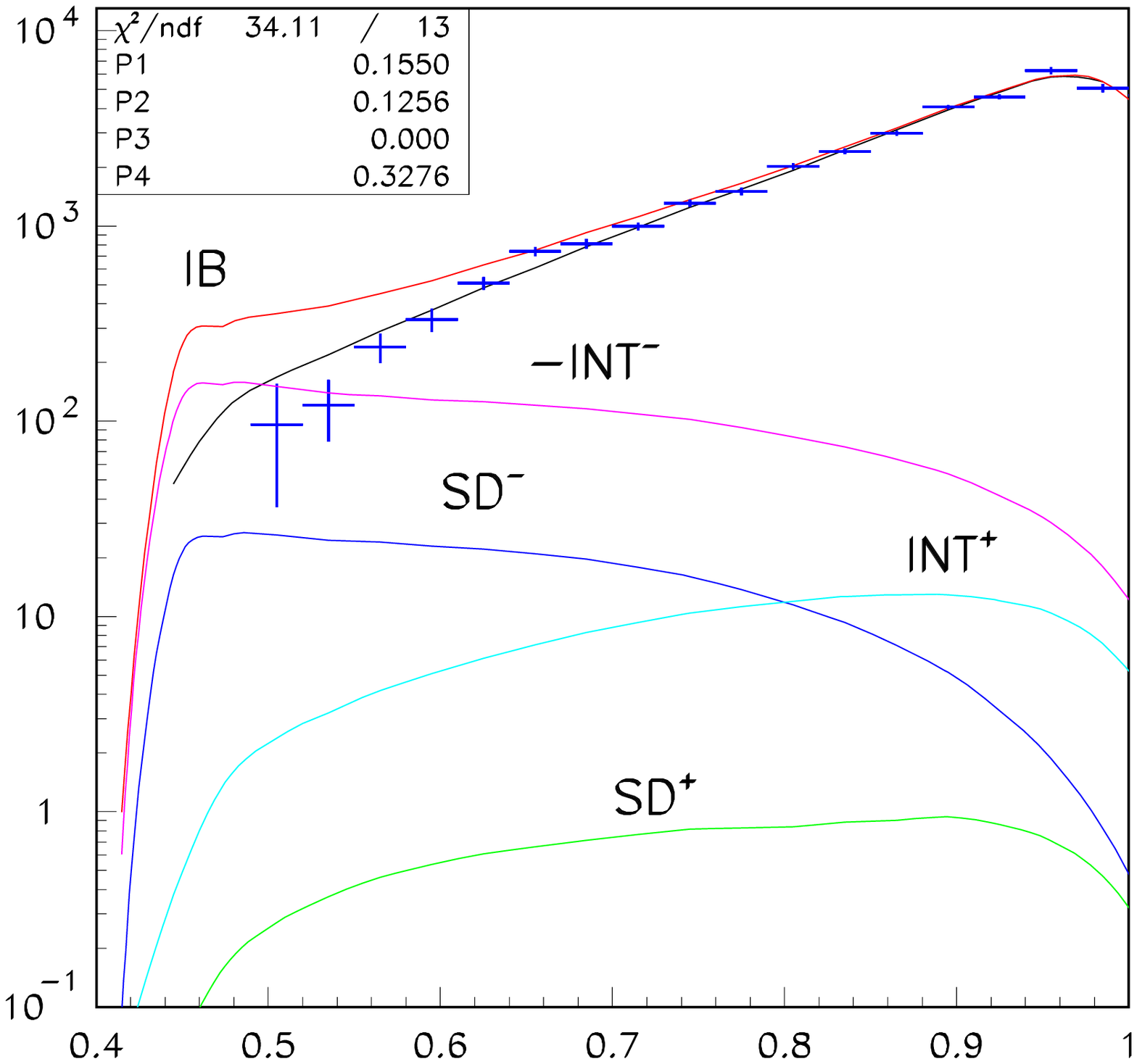,width=17.5cm}
\caption{ Results of fit of the event distribution with $F_T =0$.}
\end{figure}

  The fit is not perfect, but satisfactory, $\chi^2$/NDF$=34.11/(17-2)$ with
  around three sigma deviation from the expected $\chi^2$
   and $F_V-F_A = 0.126 \pm 0.027$. 
   
    Several sources of the systematic uncertainty have been studied. 
    The uncertainty given by poor knowledge of the background shape was
    obtained in two ways. First, by rescaling of the errors in the 
    effective mass  distributions in order to have $\chi^2$ equal to one
    in  each bin and refitting then. Second, by using polynomial
    parametrization instead of the form (10). First method leads to the
    deviation in the $F_V-F_A$ equal to 0.0104, second --- to the 0.0086,
    the maximum of these two values was used as the contribution in
    the systematic uncertainty.
    The possible contribution from the $z$ vertex position was obtained using
    increased $z$ interval with maximum $z$ equal to 1850~cm. This uncertainty
    is equal to 0.0329. The variation in the muon selection criteria gives
    the value equal to 0.0167. And the fit in the different $y$ intervals
    gives the deviation equal to 0.0191. Adding the individual errors
    in quadrature we find  a total systematic error of 0.0428.
      
  We have tried also the fit with non-zero tensor  form
  factor. The $\chi^2$/NDF$=31.97/(17-3)$,
   $F_V-F_A = 0.144 \pm 0.044 \pm 0.035$ and  
   $F_T = -0.0079 \pm 0.011 \pm 0.007$. 
  
  The fit with constant vector and axial form factors gives:
  $\chi^2$/NDF$=33.4/(17-2)$, 
  $F_V-F_A = 0.146 \pm 0.028 \pm 0.046$ with
  $F_T =0$  and $\chi^2$/NDF$=31.29/(17-3)$, 
  $F_V - F_A = 0.136 \pm 0.060 \pm 0.045$
   and $F_T = 0.006 \pm 0.035 \pm 0.007$. 

\section{ Conclusions}

  Our conclusions are as follows:    
The study of the radiative kaon decay K$^-_{\mu2\gamma}$
in a new region where SD$^-$ and INT$^-$ terms have  maximum gives the
value $F_V-F_A =0.126 \pm 0.027($stat$) \pm 0.043($syst$)$.
This value is 1.4 standard deviations above $O(p^4)$ ChPT prediction,
equal to 0.055, and probably indicates the need for higher order calculations
 or for more elaborate analysis of the $q^2$ dependence of the form
  factors.
  
  The inclusion of the tensor component gives:
 $F_V -F_A = 0.144 \pm 0.044 \pm 0.035$.
  and $F_T = -0.0079 \pm 0.011 \pm 0.007$, i.e. 
  $-0.03 < F_T < 0.01$  at 90$\%$ CL.
    
 The work 
 is  supported in part by the RFBR grant N07-02-00957(IHEP group).

\end{document}